\documentclass[12pt]{iopart}
\usepackage{graphicx}
\usepackage{cite}

\usepackage{iopams}  
\begin{document}

\title {Vortex dynamics, pinning and irreversibility field investigation in $EuRbFe_4As_4$ superconductor.}
\author{Vladimir Vlasenko$^{1}$, Kirill Pervakov$^{1}$, Sergey Gavrilkin$^{1}$} 
 \address{$^1$ P.N. Lebedev Physical Institute RAS, 119991, Moscow, Russian Federation}

\ead{vlasenkovlad@gmail.com}

\vspace{10pt}

\begin{abstract}
We performed systematic AC susceptibility and magnetic moment measurements to investigate the vortex dynamics and pinning in the $EuRbFe_4As_4$ single crystal as a function of temperature, frequency, and DC magnetic field. The vortex solid-liquid line was determined and it fits well with $H(T_p)=H_0(1-t_p)^\beta$ using $\beta$=1.74-1.91, for $H\parallel ab$. It indicates a rather high pinning strength of the vortex system. The activation energy $U_0$ was determined from thermally activated flux creep theory and reached 6700 K at low fields, suggesting strong vortex pinning. A field dependence of $U_0(H\parallel ab)\sim H^a$ with $a=0.47$ suggests thermally activated plastic pinning or caused by planar defects. Magnetic moment measurements also confirmed strong pinning in a $EuRbFe_4As_4$ superconductor and the superconducting response gives the main contribution to the $M(H)$ hysteresis. Additionally, we found evidence of long-range magnetic interactions in $Eu^{2+}$ sublattice and the FM-like nature of $Eu^{2+}$ atoms ordering. 
\end{abstract}

%
\vspace{2pc}

%
%


\section{Introduction}

Recently, a novel family of Fe-based superconductors (SC) was discovered, known as 1144-type ($AeAFe_4As_4$, where Ae = Ca,Sr,Ba,Eu and A = K,Rb,Cs) compounds \cite{Iyo}. The 1144 family exhibits intriguing and distinctive properties and has gained interest in the field of high-temperature superconductors. The self-doped $AeAFe_4As_4$  compounds possess a tetragonal structure (P4/mmm), where Ae and A layers form two inequivalent $ThCr_2Si_2$ structures \cite{Iyo}. In contrast to 122 superconductors, Fe-As bonds have a different length between the As atoms and the Fe plane. The 1144 superconducting compounds show critical temperature in range $T_c=24-36$ K \cite{Iyo,Kawashima,Liu,Kawashima2} and have estimated upper critical fields from 92 T\cite{Meier} up to 250 T \cite{Smylie}. Various transport, thermal, and thermodynamic experiments on 1144 systems show the absence of any structural transition \cite{Shiv}. However, in the case of $EuRbFe_4As_4$, bulk superconductivity with $T_c \sim 36$ K coexists with the $Eu^{2+}$ spin localization, which leads to magnetic ordering at $T_m \sim 15$ K. Therefore, $EuRbFe_4As_4$ was found to be a ferromagnetic (FM) superconductor with Eu$^{2+}$ ordering within the $ab$ plane \cite{Stolyarov, Mohammed}. However, such features can only be observed in the magnetic measurement data that indicates a spatial separation of the magnetic and superconducting regions \cite{Kawashima3}.  In contrast to Eu-1144 superconductor, the related 122-type compound ($EuFe_2As_2$) has the anti-ferromagnetic ordering which competes with the superconductivity. This exotic but distinctive behaviour in the 1144 system is encouraging the scientific community to bridge the knowledge gap in similarly unusual superconducting systems. To date, there are a little data concerning vortex dynamics, pinning and the irreversibility line behaviour in $EuRbFe_4As_4$ (Eu-1144) superconducting compounds. 
  
\section{Experimental details}

$EuRbFe_4As_4$ single crystals were grown using the self-flux method, as was done for Ni-doped Ba-122 \cite{Pervakov, Kuzmicheva}. The initial high purity components of Eu (99.95\%), Rb (99.99\%) and preliminary synthesised precursor FeAs (99.98\% Fe + 99.9999\% As) were mixed with 1:1:12 molar ratio. The mixture was placed in an alumina crucible and sealed in a niobium tube under 0.2 atm of residual argon pressure. The sealed container was loaded into a tube furnace with an argon atmosphere to suppress the alkali metal evaporation. Next, the furnace was heated up to 1250$^0C$, held at this temperature for 24 h to homogenize melting, and then cooled down to 900$^0C$ at a rate of 2$^0C/h$. At this temperature, the ampoule with crystals was held for 24 h, for growth defects elimination, and then cooled down to room temperature inside the furnace. Finally, crystals larger than 5x5 $mm^2$ in the ab-plane were collected from the crucible in an argon glove box. These crystals are shown in the inset of Fig. 1. It is clearly seen that the cleaved surface of the $EuRbFe_4As_4$ single crystal is smooth and clean. All samples were covered by parafilm to prevent oxidation and put into containers with an argon atmosphere.

Single crystal quality was examined by complementary methods. The chemical composition of the 1144 crystals was studied by a JEOL 7001F scanning electron microscope (SEM) with an INCA X-act EDS attachment. Data from several points were averaged to improve measurement accuracy. Elemental analysis in mapping mode points out a uniform distribution of the elements. We found that the chemical composition of our studied single crystal was $Eu_{1.26(2)}Rb_{0.84(2)}Fe_{4.00(5)}As_{4.01(5)}$. Excess of Eu atoms shows a presence of about 15$\%$ $EuFe_2As_2$ phase, which naturally explains stoichiometric ratio distinction.
AC susceptibility and magnetic moment measurements were performed using the Quantum Design PPMS-9 system. AC magnetic susceptibility measurements were carried out in field-cooled (FC) conditions, with $H_{ac}$ =1 $Oe$ and external magnetic field ($H\parallel ab$) up to 9 T. The frequency ($f$) of $H_{ac}$ was varied from 33 to 9777 $Hz$. DC susceptibility and M(H) measurements were done with vibrating sample magnetometer (VSM). During the DC susceptibility measurement the sample was initially cooled down to 2K in the absence of the magnetic field (ZFC procedure). At 2 K a magnetic field of 10 $Oe$ was applied, and the sample was then heated to 40 K. In the case of field cooled (FC) procedure the sample was cooled down to 2 K in the presence of 10 $Oe$ magnetic field. 

\begin{figure}[]
	\includegraphics [width=1\textwidth]{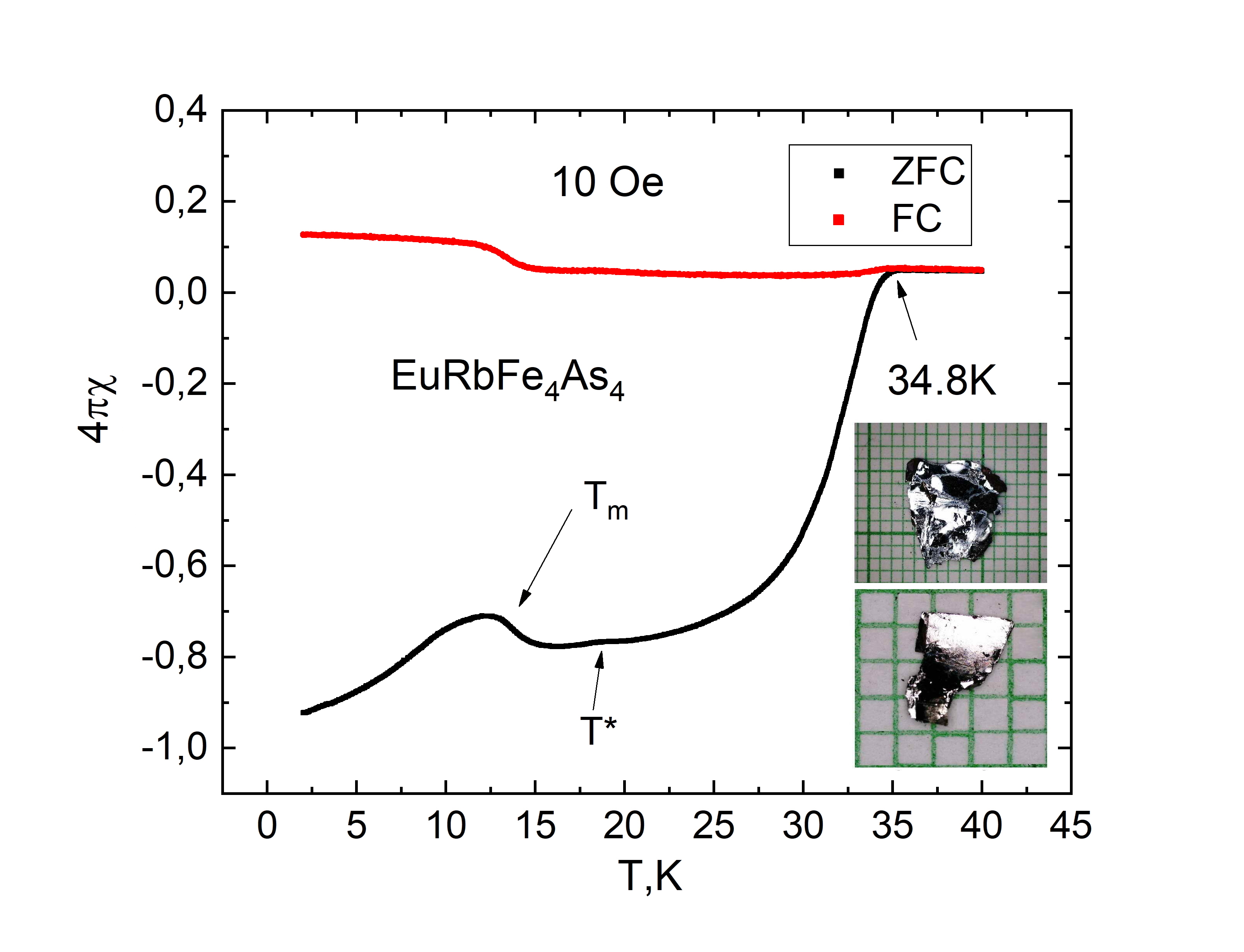}
	\caption{ (Left) Temperature dependence of the DC ZFC (zero-field cooled) and FC susceptibility in an external field of 10 Oe applied along the $ab$ axis. (Inset) Picture of the $EuRbFe_4As_4$ single crystals over a millimeter grid.}
	\label{fig:1}
\end{figure}

Fig. 1 shows the temperature dependence of DC susceptibility  for the $EuRbFe_4As_4$ single crystal along the $ab$ plane.   We define the critical temperature of the superconducting transition $T_c$ ($T_c$ = 34.8 K)  using the onset criteria. The $T_c$ of the Eu-1144 sample is slightly lower than one reported in ref. \cite{Jin}. The decrease in $T_c$ can be explained by the existence of a Rb deficiency (self-doping), which leads to extra hole-like charge carriers and induces a non-optimal doping regime in the sample \cite{Mohammed}. A magnetic feature at roughly $T_m = 14 K$ was observed and, as previously mentioned, is related to the magnetic ordering\cite{Liu2, Liu}. We also found a small sign of the Eu-122 non-superconducting phase that always accompanied 1144 phase during the crystal growth. For instance, the small feature in Fig.1 at the temperature of about 19K (T$^*$) indicates antiferromagnetic $Eu^{2+}$ ordering in Eu-122 \cite{Jiang1}. 

The initial crystal (see Fig.1)  was divided into two parts. We used smaller rectangular-shaped single crystal (S1) with dimensions of about ($a\times b \times c$) 1.12 mm, 0.65 mm, and 0.12 mm. Another part was investigated by SEM and used in other experiments.

\section{Results and discussions}

After the synthesis, we investigated stability of the crystals in the open air. For this purpose, we take crystal (S2) from an argon glove box and exposure it during 1 h in the air. It was found that $EuRbFe_4As_4$ single crystals become tarnished after air exposure. The surface is covered by a visible rainbow-coloured film. Repeated AC susceptibility measurements ($H\parallel ab$) of S2 before and after air treatment shows a reduction of the superconducting phase. The same crystal degradation was observed for other samples. Following electron microscope investigation of Eu-1144 single crystals showed clear oxidation up to 10 microns depth  in areas with an excess of rubidium on the crystal surface. Cracked exfoliated layers on the surface are observed as well. A possible explanation is that the presence of rubidium on the crystal surface leads to rubidium hydroxide formation and subsequent reactions with the underlying crystal layers. At the same time in the surface regions with a lack of rubidium or its absence, the reaction does not occur and the surface keeps smooth. The pure $EuFe_2As_2$ showed no signs of degradation (see supplementary material). 

Before investigation of the vortex pinning matter in Eu-1144, it is necessary to discuss the interaction between the magnetic and superconducting layers. To date, it has been shown that the $Eu^{2+}$ layers and superconducting condensate exhibit quasi-isolated behaviour. The facts confirming the weak interaction between the superconducting region and the $Eu^{2+}$ layers in $EuRbFe_4As_4$ are as follows \cite{Stolyarov}: i) It was shown that magnetic vortex density did not change much during the $Eu^{2+}$ ordering. ii) Magnetic ordering can only be observed within the $ab$ plane. iii) Europium replacement for non-magnetic calcium only leads to magnetic ordering suppression, but superconducting transition curves remain almost unchanged. iv) Europium atoms in the crystal structure are spatially isolated from superconducting layers.
Thus, we believe it is possible to examine the Eu-1144 superconducting state with minimal distortion induced by the $Eu^{2+}$ layers.

\begin{figure} []
	\includegraphics[width=0.7\textwidth]{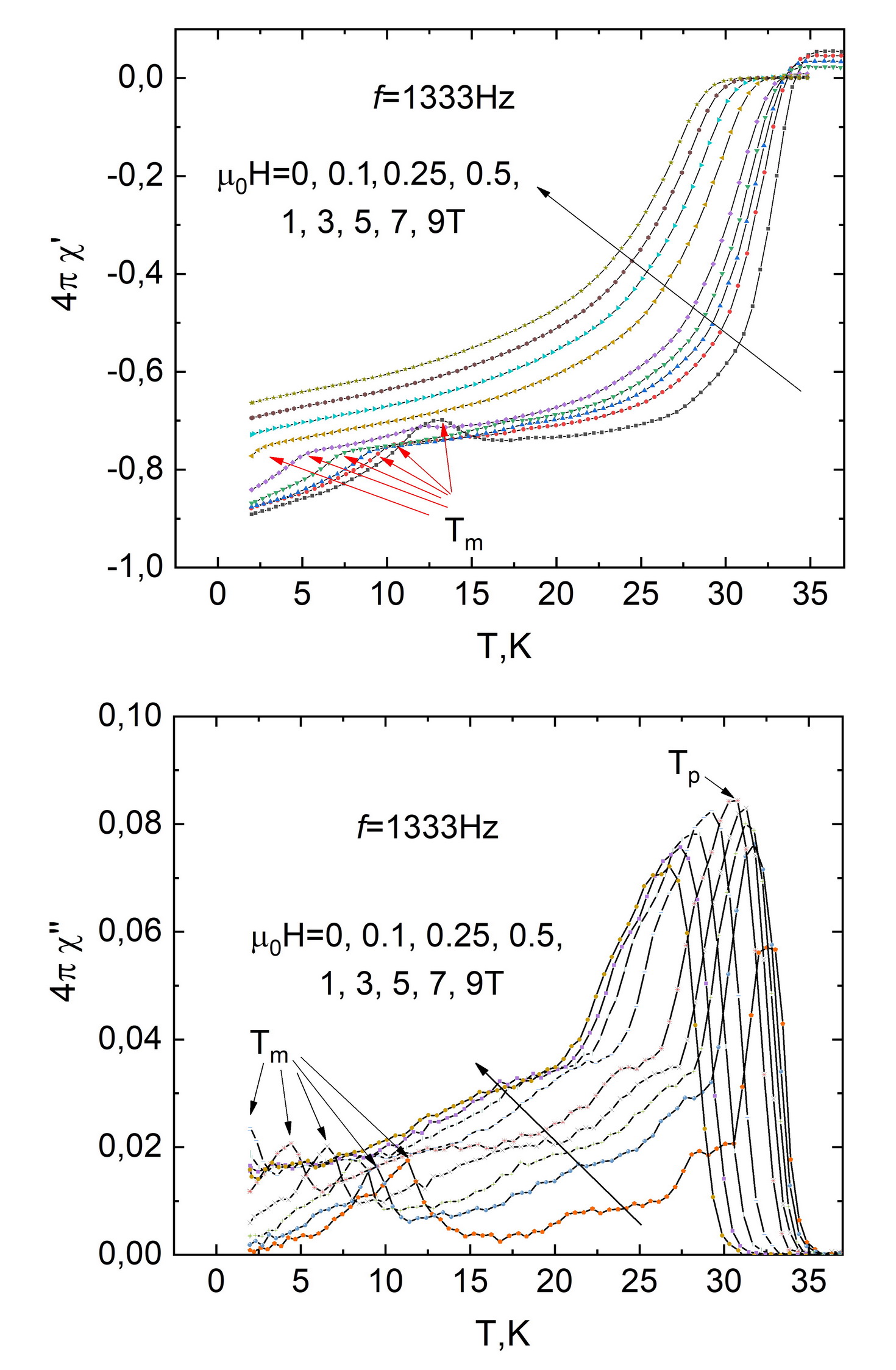}
	\caption{Temperature dependence of in-phase and out-of-phase AC susceptibility for $H\parallel ab$ at various applied magnetic fields, with $H_{ac}$ =1 Oe and frequency $f$=1333 Hz.}
	\label{fig:2}
\end{figure}

Fig.2 displays an example of  $AC$ susceptibility measurements in DC fields up to 9 T with $H_{ac}$ =1 $Oe$ at a frequency of 1333 Hz. From AC susceptibility measurements, one can found the $Eu^{2+}$ magnetic ordering has shifted from 14K to about 5K in 1T field applied. $AC$ susceptibility data at frequencies above 1333Hz show a slight kink of $Eu^{2+}$ ordering  at temperature of about 2K in 3T. Low-frequency data didn't show any features at the 3T curve. Despite our 33Hz data is quite noisy, the main features remain. For better perception we excluded plots of the 33Hz curves from following pictures but one could find the 33Hz and some other AC susceptibility measurement data in supplementary materials.
Experimental data shown in Fig. 2 were used to build $H_{с2}(T)$ and $H_{m}(T)$  phase diagram. The $H_{c2}(T)$ line was found from the $AC$ susceptibility data using the onset criteria and $H_{m}(T)$ was found as the peak position of $Eu^{2+}$ magnetic ordering. 
As shown in Fig.3 frequency-independent (f=133Hz-9777Hz) peaks are observed  in both the dispersive ($\chi'$) and dissipative ($\chi''$) parts of the AC susceptibility. Frequency-independent behaviour of the maximum on the real part of the AC susceptibility indicates a long-range magnetic interactions \cite{Dubiel1}. The presence of $\chi''$ component is the evidence of non-AFM, but FM-like ordering of the $Eu^{2+}$ atoms \cite{Balanda}. However, the plot of $\chi''$ shows that peak maximum shifts to a lower temperature (about 2K) compared to the $\chi'$ maximum. 
\begin{figure} []
	\includegraphics[width=0.9\textwidth]{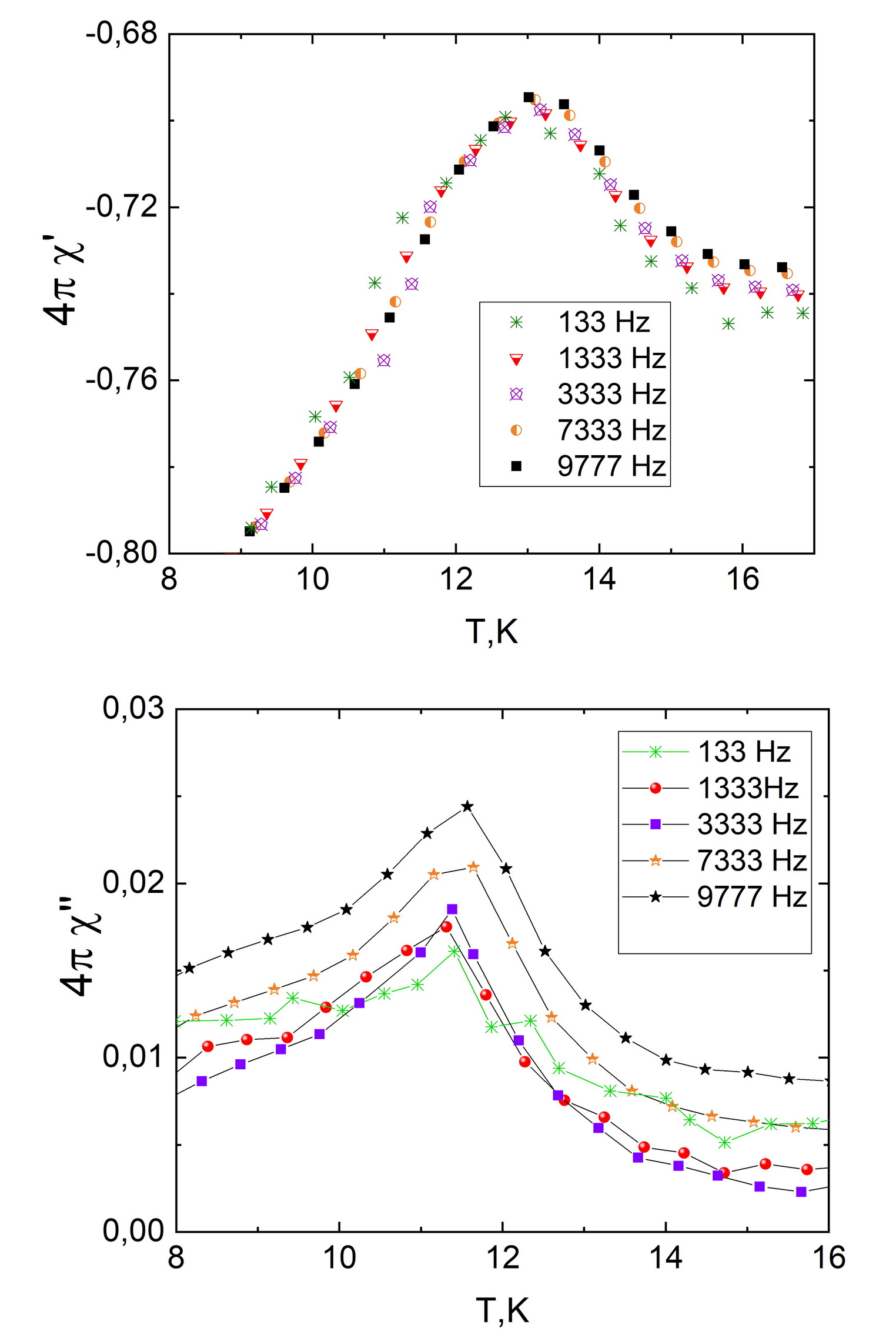}
	\caption{The frequency dependence of in-phase (real) and out-of-phase (imaginary) AC susceptibility for $H\parallel ab$ at zero magnetic fields, with $H_{ac}$ =1 Oe. }
	\label{fig:3}
\end{figure}

Vortex dynamics and pinning were also investigated by the $AC$ susceptibility measurements \cite{Prando, Prando2, Ge, Ge2}. The $AC$ susceptibility data ($\chi'$ and $\chi''$) can be interpreted using the theory of thermally activated flux motion and collective vortex pinning, based on the assumption that magnetism of Eu-1144 has weak influence on superconductivity.  
It is known that at sufficiently low frequency (and low $AC$ amplitude), the $AC$ susceptibility data could provide information about the vortex solid-to-liquid phase transition. This restriction is caused by a significant dependence of $T_p$ peak position on AC field frequency and amplitude in the $\chi''(T)$ data. When the $AC$ field frequency is lowered, $T_p$ shifts towards lower temperatures due to current density loss compensation during relaxation \cite{Ge}. In case the DC field increases, the transition in $\chi'(T)$ and $\chi''(T)$ moves to lower temperatures and the peak height increases to a certain value and remains more or less constant for further increasing $DC$ magnetic fields. 

With this consideration, we carried out systematic $AC$ susceptibility measurements, varying the measurement frequency $f$ to determine the phase transition line.  The peak  values ($T_p$) on the imaginary part ($\chi''$) of AC susceptibility were determined using the criteria detailed in ref. \cite{Prando2}. Fig. 4(a) presents $\chi''(T)$ AC susceptibility data for various frequencies near the superconducting peak region.  The $T_p$ values at various frequencies were determined and $T_p(H)$ graph is shown in Fig. 4(b). The $T_p(H)$ data can be approximated with the formula: $H(T_p)=H_0(1-t_p)^\beta$, where $t_p =T_p/T_c$ and the exponent $\beta$ provides information about the vortex pinning strength \cite{Wang,Yeshurun}. We found that in Eu-1144, $\beta$ varies from 1.74 at $f=33$ Hz to 1.91 at $f=9777$ Hz. Similar behavior was also reported for $K_{0.8}Fe_2Se_2$\cite{Ge} and $FeS_{1-x}Se_x$\cite{Wang}. In melt-textured single-domain cuprate superconductors (Gd-Ba-Cu-O) a variety of $\beta$ values have been interpreted within the framework of the flux diffusion model\cite{Ge}. However, the true nature of this phenomenon in iron-based superconductors is still under discussion. The irreversibility line ($H_{irr}(T)$) was obtained from extrapolation of $1/T_p(H)$ vs $f$ data, down to low-frequency  ln$f$ = 0 (1 Hz), as shown in Fig. 4(c) \cite{Ge}. 
 
 According to the theory, the AC susceptibility response in the vortex liquid region mainly arises from the thermally activated vortex jumps between metastable states of the vortex lattice \cite{Blatter}. In the vortex-solid state, the flux lines have to additionally overcome the pinning potential wells, and a thermally activated flux creep is observed. The main parameter that governs the vortex motion is the effective pinning barrier $U_0$, which is described, within the thermally activated framework, by the expression \cite{Buchacek,Prando2}:
\[\frac{1}{{{T}_{p}}(f )}=-\frac{1}{{{U}_{0}}(H)}\ln (\frac{f }{{{f }_{0}}}).\]

This approximation implies that $1/T_p$ depends mainly on the $U_0$ parameter, playing the role of an effective de-pinning energy barrier in the thermally activated flux creep model. Characteristic frequency parameter $f_0$ is associated with the motion of vortices (vortex hopping) around their equilibrium position ($10^{10}-10^{12}$ Hz) \cite{Blatter}.

\begin{figure} []
	\includegraphics[width=0.9\linewidth]{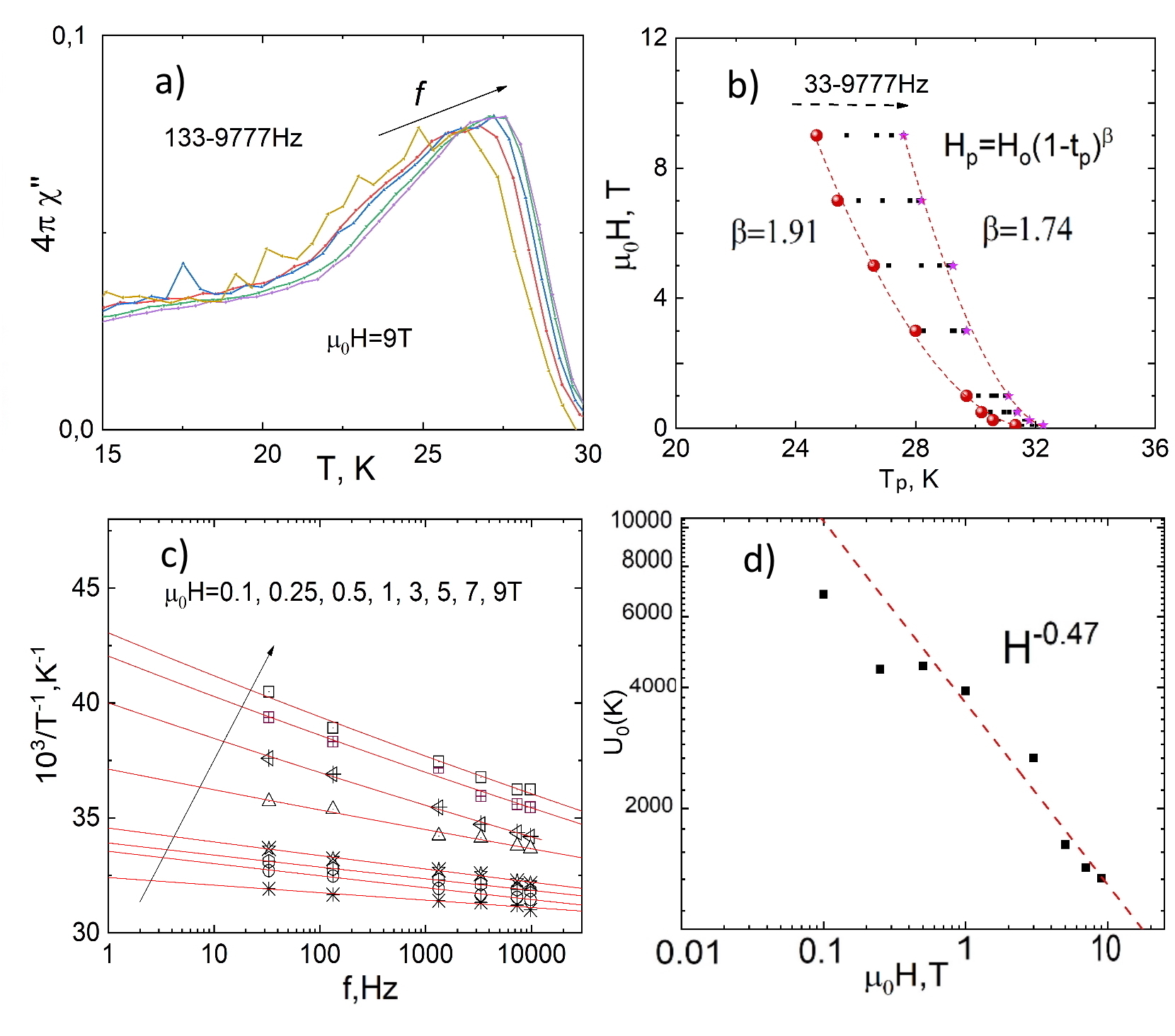}
	\caption{a) Temperature dependence of the out-of-phase AC susceptibility with H $\parallel$ ab at $H=9$ T at various frequencies. b) The $T_p$ obtained at various frequencies and magnetic fields. The dashed lines are fitted using $H(Tp)=H_0(1-t_p)^\beta$ for 33 Hz and 9777 Hz datasets, respectively. c) Logarithmic $f$ dependence of $1/T_p$ ($H_{ac}$ = 1 Oe) at various magnetic fields. The red line is the best-fit function according to Eq. 2. d) Field dependence of the activation energy $U_0(H)$.}
	\label{fig:4}
\end{figure}

In Fig. 4(d) one can observe the $U_0(H)$ data. Effective value of the pinning barrier reaches 6700 K at low fields, which is similar to reported values of other iron-based superconductors: $SmFeAsO_{0.8}F_{0.2}$ ($\sim 5\times10^4 K$)\cite{Prando}, $Ba_{0.72}K_{0.28}Fe_2As_2$ ($\sim 10^4K$) \cite{Bellingeri}, $BaFe_{1.9}Ni_{0.1}As_2$($\sim 10^4K$)\cite{Ge2}, FeS($\sim 5\times10^3K$)\cite{Wang} and $FeTe_{0.5}Se_{0.5}$ ($\sim 10^4 K$)\cite{Wang2}. The $U_0(H)$ dependence shows at least two regimes with a crossover of roughly 0.1T. Theory predicts that a low field region is associated with a single-vortex pinning regime. At higher magnetic fields a power-law dependence $U_0 \propto H^{-a}$ may be observed. We found $U_0(H)$ follows an $H^{-0.47}$ dependence at magnetic fields above 0.1T. However, collective pinning with $U_0 \sim H^{-1}$ is not the case. Another possible pinning mechanism here, considering $U_0 \sim H^{-0.47}$ at $H>0.1T$, is might be flux pinning due to the planar defects \cite{BARTOLOME}. According to very recent TEM observations in CaK-1144 the presence of planar defects along the $ab$ plane was found \cite{Ishida}. Given a similar synthesis method (FeAs flux), our EDX data and the same structure of $EuRbFe_4As_4$, it can be assumed that in our case $EuFe_2As_2$ layers will act as planar defects. However, unlikely to CaK-1144, in Eu-1144  intergrowths are regarded as non-SC planar defects. This fact explains features absence in $J_c\parallel ab$ curves  at low fields compared to CaK-1144 \cite{Ishida}. 

 \begin{figure} []
 	\includegraphics[width=0.7\textwidth]{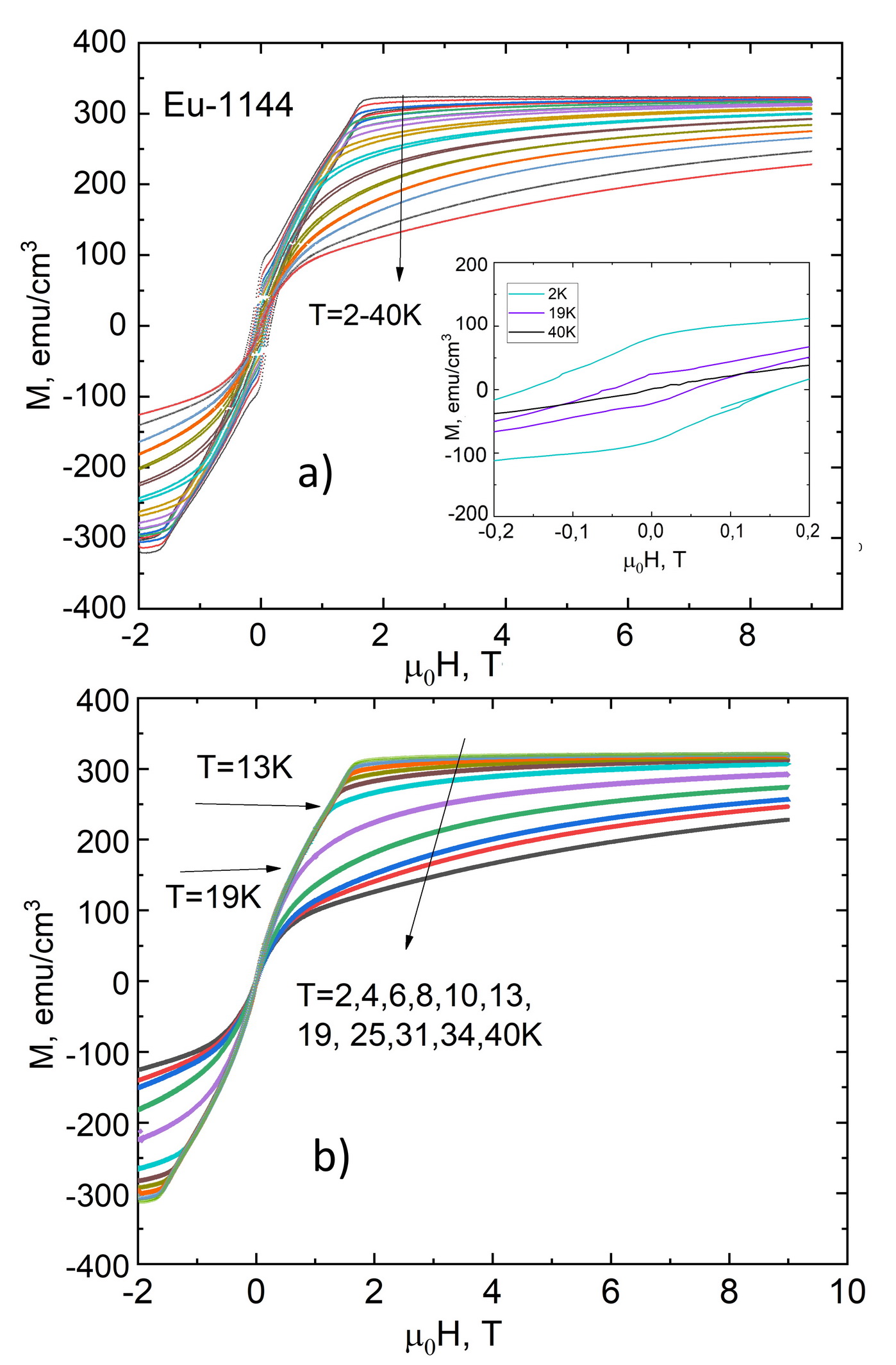}
 	\caption{(a)Isothermal magnetization hysteresis loops for Eu-1144, as a function of magnetic field (up to 9 T) with $H\parallel ab$. (Inset) Low field magnetization loops for temperatures above and below $T_c$. (b) Calculated magnetization curve of the $Eu^{2+}$ sub-lattice of the Eu-1144 superconducting sample.}
 	\label{fig:5}
 \end{figure}

We also investigated Eu-1144 with isothermal magnetization measurements M(H), presented in Fig.5a. All magnetization hysteresis loops M(H, T) data were collected with the magnetic field parallel to the $ab$ axis, the sweep rate was 100 Oe/s. In the inset of Fig.5a the magnetization loops at 2, 19 and 40K at low fields are shown. One can see that magnetization measurement above the $T_c$ shows no hysteresis, however, the Brillouin-function-like curve is observed at 40 K. This non-linear  behaviour of  M(H) at 40K compared to ref.\cite{Smylie} might be due to the presence of the $EuFe_2As_2$ in the studied sample. At 19 K the M(H) loop originates from the superconductivity of $EuRbFe_4As_4$. Also the observed distortion is related to the paramagnetism. At lower  temperatures ($T<14K$), the M(H) loops look like a combination of a ferromagnetic-like \cite{Mohammed} and superconducting response due to the flux pinning. 
Assuming that the $Eu^{2+}$ signal is small compared to the superconducting hysteresis and M(H) curves are symmetric, we can extract the magnetization curve of the $Eu^{2+}$ sublattice using a simple formula: $M_{Eu}=(M_++M_-)/2$, where $M_+$ and $M_-$ is the magnetization measured in increasing and decreasing applied field, respectively \cite{Smylie}. The results are shown in Fig.5b, whereas magnetization curves saturate at about 310 $emu/cm^3$ at 2K, corresponding to 6.4$\mu_B$/Eu. The obtained value is slightly less than the calculated full moment of 7$\mu_B$/Eu for ferromagnetic alignment. The above results also indicate an FM-like transition at about $T_m$ = 14 K for $EuRbFe_4As_4$ compound.
 It should be noted that the $EuFe_2As_2$ phase should not give a significant change in the overall magnetic moment of the $Eu^{2+}$ atoms, because the cell volume (184.5 vs. 201 \AA$^3$\cite{Uhoia,Kawashima3}), number of $Eu^{2+}$ atoms per unit cell and magnetic moment along $ab$ plane at low temperatures \cite{Jiang1} are almost the same compared to Eu-1144.

In general, there is a complex interplay between magnetism and superconductivity in magnetic superconductors. It makes difficult to produce a quantitative estimate of $J_c$. However, the qualitative aspect remains \cite{Vlasko}. Therefore, we plot $J_c(H)$ curves at different temperatures using the Bean’s critical state model \cite{Bean} in Fig.6a. According to theoretical predictions $J_c\sim {\Delta} M$, where $\Delta M$ is the hysteresis loop width. At temperatures above 20K there are no magnetic transitions, which may give significant contribution to the M(H) loop width, and therefore to the $\Delta M$ value. The $EuFe_2As_2$ ferromagnetic ordering at about 19K is rather small and saturates at about 1T (T=2K) in field orientation along the $ab$ plane \cite{Jiang1}. In Fig.6a $J_c(H)$ curve at 2K does not show significant features at about 1T applied magnetic field. The same situation is observed for the Eu-1144 compound. In case $Eu^{2+}$ magnetic ordering gives a considerable magnetic hysteresis, $\Delta M(H)$ data should show a noticeable field features below 1.5T at 2K. However, there are no considerable peculiarities on the $\Delta M(H)$ curve at H$<$1.5T. Moreover, in the inset of Fig.6a the normalized $J_c(T)$ ($H \parallel ab$)  data taken at 0T did not show any changes up to 8K, and the $J_c/J_c (0)$ vs. $T/T_c$ curve in Eu-1144 is similar to one for the 122 system without a magnetic transition  \cite{Tanatar}. The $J_c(0)$ was evaluated by $J_c(T)$ data fitting with formula $J_c=Jc(0)*(1-(T/T_c)^n)^m$ \cite{Vlasenko}, where $n$ and $m$ are  parameters. Given that the ${\Delta} M$ width did not vastly change above or below 15 K, we could conclude that the superconducting response is the main contribution to the $M(H)$ hysteresis. Thus, the influence of $Eu^{2+}$ magnetic ordering on the ${\Delta}M$ values  is insignificant compared to SC signal even at low temperatures. Our results are consistent with the assumption presented in ref.\cite{Smylie}.

\begin{figure} []
	\includegraphics[width=0.9\textwidth]{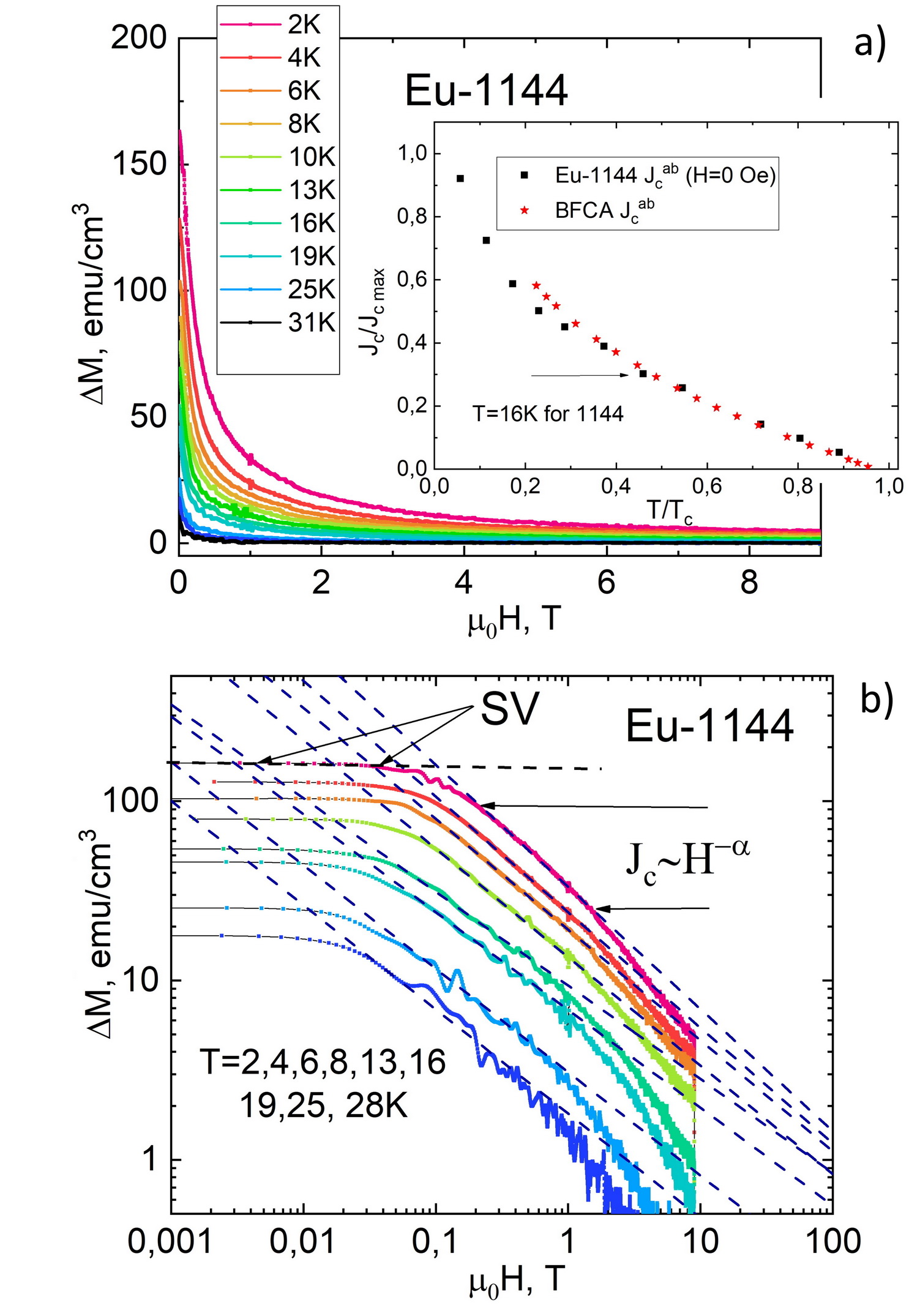}
	\caption{ (a) Field dependence of $J_c$ $\propto$ $\Delta M$ in the  H $\parallel$ ab for Eu-1144 sample. Inset: The Eu-1144 and Ba-122 \cite{Tanatar} normalized critical current $J_c/J_c(0)$ vs. normalized critical temperature ($T/T_c$) plot. The arrow shows the point just before $Eu^{2+}$ magnetic ordering. (b) The ${\Delta}M$ curve is plotted in a double logarithmic scale for selected temperatures. The dash lines show the power law plot $B^{-n}$, where $0.55<n<0.65$ is expected for the strong pinning.}
\end{figure}

In order to clarify the pinning nature in Eu-1144, we plot ${\Delta}M$ versus $H$ with log-log scale at different temperatures, the result is shown in Fig. 6 (Bottom). At low fields, typically about $100-350$ Oe, $J_c$ is independent on external applied magnetic field - single vortex (SV) regime is observed. At higher magnetic fields, from 0.1-0.2T up to 1.5 T, the critical current follows a power law behavior $J_c$ $\propto$ $H^{-a}$ with $0.55<a<0.68$. The $a$ exponent values obtained in this work are in good agreement with the theoretical prediction of H$^{-5/8}$, which indicates strong vortex pinning \cite{Beek2}. Further increase of the magnetic field leads to a rapid decrease in $J_c$ related to changes in the vortex dynamics\cite{Haberkorn,Haberkorn2}.

To summarize our data we built the magnetic phase diagram of $EuRbFe_4As_4$ as shown in Fig. 7.  We have plotted temperature dependences of the characteristic fields such as $H_{c2}$, $H_{irr}$  and $H_m$. Above the upper critical field ($H_{c2}$) $EuRbFe_4As_4$ single crystal  is in a normal state (NS). Below the $H_{c2}(T)$ line vortices are in an unpinned liquid state (VL). The irreversibility line $H_{irr}$ separates a solid (pinned) vortex lattice/glass (VS) and vortex liquid state. 

\begin{figure} []
	\includegraphics[width=1\textwidth]{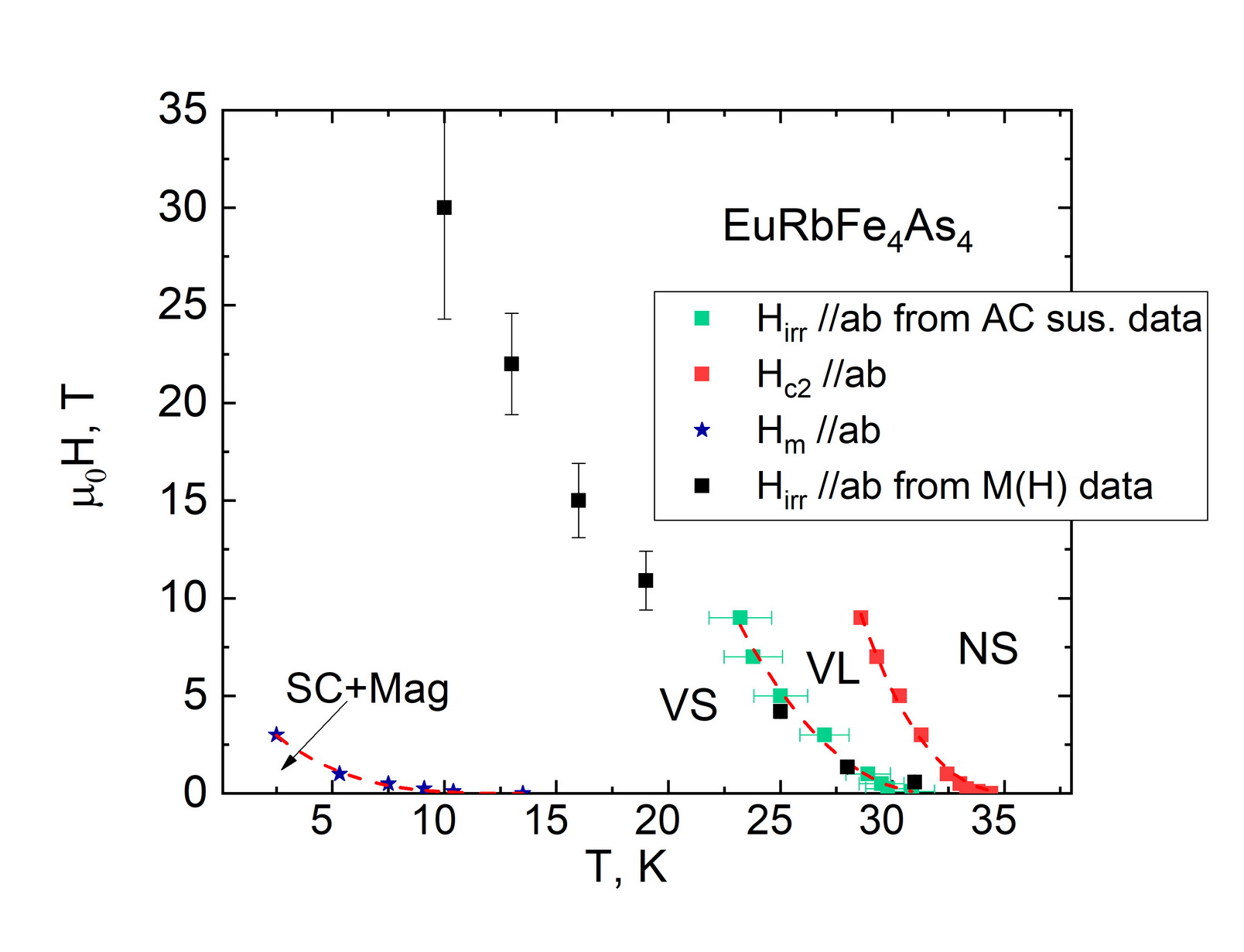}
	\caption{Vortex phase diagram of the $EuRbFe_4As_4$ single
		crystal. The dashed lines are plots using the empirical formula $H(T) = H(0)[1-T/T_c]^n$ yielding $n$ = 2.15, 2.05, and 3.48 for $H_{irr}$, $H_{c2}$, and $H_m$. The $NS$, $VL$, $VS$ and $SC+M$ are normal state, vortex liquid, vortex solid and coexistence of superconductivity with $Eu^{2+}$ magnetic ordering, respectively. }
\end{figure}

Temperature dependence of $H_m$ marks the crossover to a superconductor with magnetic ordered $Eu^{2+}$ layers (SC+M).
We also estimate the $H_{irr}(T)$ values from ${\Delta}M$ data. One can see that in Fig. 5 the ${\Delta}M(H)$ curves  have a similar shape. Assuming  the scaling of ${\Delta}M(H)$ vs. $T$ we estimate $H_{irr}$ using $\approx1 emu/g$ criteria up to 10K. At lower temperatures the  $H_{irr}$ value increase is less than the increment of the estimation error. The temperature dependence of $H_{c2}$, $H_{irr}$ and  $H_m$  can be fitted well  by the simple functional form $H(T) = H(0)(1-T/T_c)^n$, where $n$ is an exponent. From the interpolation of our experimental data, we obtain $n$ = 1.38,  $H_{irr}(0)$ = 70 T and $n$ = 2.15, $H_{c2}$=135T. The obtained values of exponent $n$ are similar to  other Fe-based superconductors\cite{Pramanik,Shen2}. 
It is well known that a rough estimation of the $H_{c2}(0)$ values also can be obtained from  $dH_{c2}/dT$ slopes near $T_c$, using the Werthamer-Helfand-Hohenberg (WHH) formula. For single band SC ${{H}_{c2}}(0)=-0.693 dH/dT{{|}_{T_c}}\times {{T}_{c}}$ \cite{WHH}. However, the WHH model as it was shown in direct high-field measurements gives an overestimated value for 1144 superconducting family, especially at H//ab. Very recent, direct high-field measurements in a $CaNaFe_4As_4$ superconductor (with the similar $T_c$ and structure), $T_c=35 K$,  $dH^{ab}_{c2}/dT$ $\sim$ $-5 T/K^{-1}$ and $dH^{c}_{c2}/dT$  $\sim$ $-10 T/K^{-1}$ show that upper critical field, in both cases, is less or about 100T \cite{Meier}. Therefore $H^{ab}_{c2}(0)$ real values should be significantly lower than estimated from WHH 150T \cite{Kawashima} and 250T \cite{Smylie}. Eu-1144 with the same structure and critical temperature  should have similar value of upper critical field. This was very recently demonstrated in direct high-field (up to 65T) measurements\cite{Smylie3}. Authors found the $H_{c2}$ values are in the range of 70-80 T. It was also concluded that the Pauli limit prevails for $H\parallel ab$, but orbital limitation is more prevalent for $H\parallel c$. Moreover, it was assumed that the 1144 compounds are a possible candidate for the existence of the FFLO state \cite{Smylie3}.

\section{Conclusions}
In summary, detailed AC susceptibility and magnetic moment measurements were performed to study the vortex pinning and flux dynamics in $EuRbFe_4As_4$ single crystals. We found evidence of long-range magnetic interactions in $Eu^{2+}$ sublattice and possible the FM-like nature of $Eu^{2+}$ atoms ordering.
Our estimation of the activation energy from AC susceptibility data give us value about 6700 K at low fields. In DC fields higher than 0.1 T, the $U_0$ decays with $H^{-0.47}$, which  is typical for thermally activated vortex plastic motion. However, considering the layered structure of this material, with $EuFe_2As_2$ non-SC intergrowths, flux pinning may be caused by planar defects. The $J_c$$\sim$ $\Delta$M(H) data evidenced to strong vortex pinning in Eu-1144. Additionally, we showed that $Eu^{2+}$ magnetic ordering is shifted to 5K at fields about 1T. To summarize the experimental data, we built the magnetic phase diagram of the $EuRbFe_4As_4$ superconductor and the upper critical field, irreversibility and Eu$^{2+}$ magnetic ordering lines were found. 


\section{Acknowledgments}
 The measurements were done using research equipment of the Shared Facilities Center at LPI.
\\
\section{References.\label{bibby}}


\begin{thebibliography}{}
	
\bibitem{Iyo}
	 Iyo A et al 2016 
 {\it J. Am.Chem. Soc.} {\bf 138} 3410
	
	\bibitem{Kawashima}
	Kawashima  K et al 2016
	 {\it J. Phys.Soc. Jpn.} {\bf 85} 064710
	
	\bibitem{Liu}
	Liu Y et al 2016 {\it Phys.Rev.} B {\bf 93} 214503 
	
	\bibitem{Kawashima2}
	Kawashima K et al 2018
	{\it J. Phys. Chem. Lett.}, {\bf 9} 868-873 
	
	\bibitem{Meier}
	Meier W R et al 2016 {\it Phys. Rev.} B {\bf 94} 064501
	
	\bibitem{Smylie}
	Smylie M P et al 2018  {\it Phys. Rev.} B {\bf 98} 104503

\bibitem{Shiv}
	Singh S J et al 2018 {\it Phys. Rev. Mat.} {\bf 2} 074802

\bibitem{Mohammed}
	 Albedah M A, Nejadsattari F , Stadnik Z M,  Liu Yi, and  Cao Guang-Han 2018 {\it Phys. Rev.} B {\bf 97} 144426 

\bibitem{Stolyarov}
	Stolyarov V S et al 2018 {\it Phys. Rev.} B {\bf 98} 140506(R) 
	
	\bibitem{Kawashima3}
	Kawashima K et al 2018 {\it J. Phys.: Conf. Ser.} {\bf 969} 012027
	
		\bibitem{Pervakov}
	Pervakov K S,  Vlasenko V A,  Khlybov E P,  Zaleski A,  Pudalov V M and  Eltsev Yu F 2013 {\it Supercond. Sci. Technol.} {\bf 26}  015008 
	
	\bibitem{Kuzmicheva}
	Kuzmicheva T E et al 2017 {\it Phys. Usp.} {\bf 60} 419–429
	
	\bibitem{Jin}
	Bao Jin-Ke et al 2018 {\it Cryst. Growth Des.}   {\bf 18} 6 3517-3523 
	
	\bibitem{Liu2}
	Liu Yi,  Liu Ya-Bin, Yu Ya-Long, Tao Qian, Feng Chun-Mu  and  Cao Guang-Han 2017 {\it Phys. Rev.} B {\bf 96} 224510
	
	\bibitem{Jiang1}
	Jiang S et al 2009 {\it New J. Phys.} {\bf 11}  025007	
	
	\bibitem{Dubiel1}
	Dubiel S M, Tsindlekht  M I, Felner I 2015 {\it Journal of Alloys and Compounds} {\bf 642} 177–179
	
	\bibitem{Balanda}
	Balanda M 2013 {\it Acta Phys. Pol.  } A {\bf 124} (6) 964-976
		
	\bibitem{Prando}
	Prando G et al 2011 {\it Phys. Rev.} B {\bf 83} 174514
	
	\bibitem{Prando2}
	Prando G et al 2012 {\it Phys. Rev.} B {\bf 85} 144522
	
	\bibitem{Ge}
	Ge J, Gutierrez J, Li M, Zhang J and  Moshchalkov V V 2013 {\it Appl. Phys. Lett.} {\bf 103} 052602	
	
	\bibitem{Ge2}
	Ge Jun-Yi,  Li Lin-Jun, Xu Zhu-An and Moshchalkov V V 2016 {\it Journal of Applied Physics} {\bf 119}, 163904
	
		\bibitem{Yeshurun}
	Yeshurun Y and Malozemoff A P  1988 {\it Phys. Rev. Lett.} {\bf 60} 2202
	
		\bibitem{Wang}
	Wang A and Petrovic C 2017 {\it Appl. Phys. Lett.} {\bf 110} 232601
	
		\bibitem{Blatter}
	Blatter G, Feigel’man M V,   Geshkenbein V B, Larkin A I,  Vinokur V M  1994 {\it Rev. Mod. Phys.} {\bf 66} 11–25
	
	\bibitem{Buchacek}
	Buchacek M, Willa R, Geshkenbein V B and Blatter G 2018 {\it Phys. Rev.} B {\bf 98} 094510
	
		\bibitem{Bellingeri}
	Bellingeri E {\it et al} 2012 {\it Appl. Phys. Lett.} {\bf 100} 082601
	
	
	\bibitem{Wang2}
	Wang X-L {\it et al} 2010 {\it Phys. Rev.} B {\bf 82} 024525
	
		\bibitem{BARTOLOME}
	Bartolome E, Palau A, Llordes A, Puig T and Obradors X 2010 {\it Phys. Rev.} B {\bf 81} 184530
	
	\bibitem{Ishida}
    Ishida S et al. 2019 {\it npj Quantum Materials}  {\bf 4} 27
	
	\bibitem{Uhoia}
	Uhoya W et al. 2010 {\it J. Phys.: Condens. Matter}  {\bf 22} 292202 (5pp)
	
	\bibitem{Vlasko}
	Vlasko-Vlasov V K et al 2019 {\it Phys. Rev.} B {\bf 99} 134503  (Supplementary material)


	\bibitem{Bean}
	 Bean C P 1964 {\it Rev. Mod. Phys.} {\bf 36} 31 1964
	 
	 \bibitem{Tanatar}
	 Tanatar et al. 2009 {\it Phys. Rev.} B {\bf 79} 094507
	 
	 \bibitem{Vlasenko}
	 Vlasenko V A, Pervakov K S,  Gavrilkin  S Yu, Eltsev Yu F 2015 {\it Physics Procedia} {\bf 67} 952 – 957 
	
	
	\bibitem{Haberkorn}
	Haberkorn N, Miura M, Maiorov B, Chen  G F, Yu W and  Civale L 2011 {\it  Phys. Rev.} B {\bf 84} 094522 
	
	\bibitem{Haberkorn2}
	Haberkorn N et al 2012 {\it  Phys. Rev.} B {\bf 85} 014522 
	
	\bibitem{Beek2}
	 Beek C J et al 2010 {\it Phys. Rev.} B {\bf 81} 174517 
	
	\bibitem{Pramanik} 
	Pramanik A K, Aswartham  S, Wolter A U B, Wurmehl S, Kataev V and Buchner B, 2013 {\it J. Phys.: Condens. Matter} {\bf 25} 495701 
	
	\bibitem{Shen2} 
	Shen B, Cheng P,  Wang Z, Fang L, Ren C, Shan L, and  Wen H-H 2010 {\it Phys. Rev.} B {\bf 81} 014503
	
	\bibitem{WHH}
	Werthamer N R, Helfand E and Hohenberg P C  1966 {\it Phys. Rev.} {\bf 147} 295
	
	\bibitem{Khim}
	Khim S, Lee B, Kim J W, Choi E S, Stewart G R and  Kim K H 2011 {\it Phys. Rev.} B {\bf 84} 104502
	
	\bibitem{Smylie3}
	Smylie M P {\it et al} 2019 {\it Phys. Rev.}  B {\bf 100} 054507 
	
	
\end{thebibliography}
\end{document}